\documentclass{aastex}
\usepackage{spr-astr-addons}

\begin{document}

\title{Angular momentum distribution during the collapse of 
primordial star-forming clouds}

\shorttitle{Angular momentum profile}
\shortauthors{Jayanta Dutta}

\author{Jayanta Dutta}
\affil{$^1$Instituto de Astrofísica e Ciências do Espaço, Universidade de Lisboa, OAL, Tapada da Ajuda, PT1349-018 Lisboa, Portugal\\
$^2$ Raman Research Institute, C.~V.~Raman Avenue, Sadashivanagar, 
Bengaluru, Karnataka 560080\\ 
jd.astrop@gmail.com}

\begin{abstract}
It {\bf is generally believed} that angular momentum is distributed 
during the gravitational collapse of the primordial star forming 
cloud. However, so far there has been little understanding of 
the exact details of the distribution. We use the modified 
version of the Gadget-2 code, a three\hbox{-}dimensional 
smoothed\hbox{-}particle hydrodynamics {\bf simulation}, to follow 
the evolution of the collapsing gas in both idealized as well 
as more realistic minihalos. We find that, despite the lack of 
any initial turbulence and magnetic fields in the clouds the 
angular momentum profile follows the same characteristic 
power\hbox{-}law that has been reported in studies that employed 
fully self\hbox{-}consistent cosmological initial conditions. 
The fit of the power\hbox{-}law appears to be roughly constant 
regardless of the initial rotation of the cloud. We conclude 
that the specific angular momentum of the self-gravitating 
rotating gas in the primordial minihalos maintains a scaling 
relation with the gas mass as $L \propto M^{1.125}$. We also
discuss the plausible mechanisms for the power-law distribution.
\end{abstract}


\maketitle

\section{Introduction}
\label{sec:introduction}

The significant progress in the theoretical as well as numerical 
calculations {\bf has provided a} great stride in our understanding 
of the history of the Universe, especially the transition period 
from the `Cosmic Dark Ages' to the first source of light to the 
present-day clumpy structure of the Universe 
\citep{madao2000,bl01,hw02,un03,sfsob03,frebel05}.
In last two decades, there have been extensive studies on the 
hierarchical structure formation that eventually leads to the collapse 
of the first gravitationally and thermally unstable primordial gas 
clouds 
\citep{reed05,cf05,yoha06,on07}. 
{\bf It has been shown that the angular momentum of the dark matter 
(DM) and baryons become comparable during virialization \citep{wa07}}.  
The physics that determines the collapse is extremely complex along 
with {\bf a complicated} chemical network, and is intrinsically 
multidimensional, nonlinear and {\bf not in equilibrium}
\citep{ga08,whalen08,byhm09,hoyy11,johnson13}.

One of the interesting physical problems {\bf is the distribution of the
angular momentum of the baryons that have collapsed gravitationally
within minihalos \citep{larson69,larson72,silk77,norman80,cr86}. This
process} ultimately leads to the formation and evolution of the 
very first stars in the Universe, the so-called Population~III (or 
Pop~III) stars (e.g.\ \citealt{htl96,tegmark97,on98}). {\bf Numerical 
computations using various methods (primarily Lagrangian and Eulerian 
code) make it possible to study the gas dynamics with higher spatial 
resolution} \cite[e.g.,][]{bcl99,abn00}. These calculations conclude 
that the angular momentum of the continually collapsing core is just 
a fraction of the total angular momentum of the cloud in which the 
high-density core forms \citep{abn02,fisher04,jap04}. Clearly 
angular momentum must be redistributed during the star formation 
\citep{on08,yoh08,tao09}. 

{\bf The angular momentum in protogalaxies also has substantial 
effect on the temperature evolution \citep{vvs2010}. They concluded 
that the increase of rotational angular momentum of the gas results
in cooling of the gas to a temperature less than 100 K. It is therefore
obvious that the existence of initial angular momentum can delay the 
formation of the primordial stars. Even the efficiency of fragmentation
and the star formation rate in primordial gas becomes sensitive to 
the angular momentum \citep{bcl02}}.
{\bf The same conclusion has been drawn by \citet{dutta12} that shows 
that the cloud with higher angular momentum leads the gas to have 
a lower temperature compared to their slowly rotating counterparts.}
More recently, 
the angular speed of the first stars has been investigated thoroughly 
using the sink particles method {\bf (a computational technique in 
which a certain high density regime is} considered as a growing 
protostars that can accrete gas particles) in cosmological simulations 
by \citet{sbl11}. Although angular momentum is conserved during the 
collapse of a rotating gaseous cloud, a small fraction of the cloud's 
mass at the center can collapse to stellar densities, leaving behind 
the rest of cloud in an extended rotationally supported envelope that 
fragments \citep{cgsgkb11}. The fraction of the collapsing cloud that 
ends up in a star or stellar system depends on the amount of angular 
momentum transferred outward during the collapse \cite[e.g.,][]
{gbcgskys12,sb13}.

{\bf The recent simulations, e.g., using the moving mesh code 
{\em Arepo} by \citet{greif15} and using smoothed particle 
hydrodynamics (SPH) by \citet{dutta15b}} have addressed the important 
issues {\bf such as} transfer and redistribution of the angular 
momentum while describing the velocity structure, accretion and 
thermal evolution of gas in a number of minihalos. The analytic 
models for the rapidly rotating disks in the presence of turbulence 
and UV backgrounds have been discussed in a series of papers 
\cite[][and references therein]{latif15}. In addition, there is 
strong indication of the dependency of the angular momentum on the 
Pop~III accretion rate in the current literature \citep{hirano15}. 
However, so far there has been little understanding on the {\bf basic
characteristic of} cloud's angular momentum distribution and its 
evolution during the gravitational collapse of the primordial star 
forming gas. The extent to which the angular momentum of the cloud 
depends on the initial strength of rotation has never been 
systematically tested. {\bf Moreover, there is still ambiguity of 
the basic physical mechanism that is responsible for the distribution
of angular momentum in primordial clouds.} In our recent study,
we have {\bf taken care to address} the influence of the initial 
rotation of the primordial cloud on the thermal, chemical and 
dynamical evolution that results in the fragmentation of the 
disk \citep{dutta15b}. In this work, we revisit the physics of 
gravitational collapse of the primordial gas with a particular 
focus on the angular momentum profile in the evolutionary stage. 
Our approach to investigate the gas evolution is different from 
the previous studies in the sense that we {\bf scrutinize} the 
angular momentum distribution for a large set of simulations with 
a varied initial degree of rotation of the gas clump {\bf and 
compare the result with more realistic cosmological simulations
that was based on different numerical formalism}. We have studied 
the gas collapse till it attains the protostellar densities. 

In the next section \S \ref{sec:gadget2}, we briefly outline 
the numerical set-up of simulations and initial conditions. 
We then discuss the relevant physical concept of the problem 
with an emphasis on the origin and distribution of the angular 
momentum of the collapsing gas in \S \ref{sec:angular}. We 
draw our conclusions in \S \ref{sec:summary}.

%
%
%
%
%
%

\section{Simulations}
\label{sec:gadget2}

To investigate the angular momentum distribution, we use a simple 
and straightforward numerical set-up where the gravitationally
unstable gas clump is assumed to be spherical of size $\sim 2.7$ pc.
For simplicity, we assume that gas particles are uniformly distributed 
with an initial number density $n_0 = 10^{3}$ cm$^{-3}$ and temperature
$T_0 = 200$ K. These values are consistent with the primordial gas 
cloud at redshift $z \geq 20$ \citep{abn02,bl04}.

{\bf Initially the gas clump contains only atomic hydrogen that will 
collapse only if the gravitational potential energy of the cloud
is greater than the net kinetic energy. With no external source
of cooling, the above criteria is fulfilled because the free-fall 
time ($t_{\rm ff} = \sqrt{3\pi/32G\rho} = 1.37$ Myr) is shorter 
than the sound-crossing time $t_{\rm sc} \approx $ 5 Myr. The gas 
cloud therefore inevitably starts to collapse under the self-gravity.}
We use 5 million SPH particles to model the gas distribution in the 
cloud of total mass of $M \sim 3000$ $M_\odot$. The numerical 
resolution (for 100 SPH particles) is $0.06$ $M_\odot$.

With this set-up, the artificial clump is given initial angular 
momentum that can be obtained from the rotational ($E_{\rm rot}$) 
and the gravitational ($E_{\rm grav}$) energies of gas particles. 
We follow \citet{sdz03} to model the angular velocity of the cloud 
in which the decisive parameter $\beta_0$, where 
$\beta_0 = E_{\rm rot}/E_{\rm grav} = R_0^3\Omega_0^2/3GM_0$, 
sets the strength of rotational support of the gas clump. Here $R_0$ 
and $\Omega_0$ are the radius and angular velocity of the gas clump. 
We perform six numerical experiments with varied $\beta_0$ parameters, 
{\bf whose magnitude spans two orders of magnitude.}

We take care to ensure that our investigation is not biased {\bf by} the 
artificial configuration of the gas clump. Hence we investigate the 
angular momentum distribution in two different minihalos (CH1 and 
CH2 of \citealt{dnck15}) obtained from cosmological simulations of 
the hydrodynamic moving mesh code {\em Arepo} \citep{springel10}.
We implement these realistic minihalos in our modified SPH Gadget-2 
code \citep{springel05}. {\bf We here mention that although both 
the codes, i.e., {\em Arepo} and Gadget-2 use different approach in 
solving hydrodynamics equations, both of them basically follow the 
Lagrangian formalism. This enables us to use the minihalos from the
unstructured moving mesh into SPH configuration.} 

We summarise here the detailed configuration and initial conditions 
of these minihalos (see table~1 of \citealt{dutta15b} for the 
detailed configuration of the halos). 
{\bf The gas density of both minihalos increases towards the centre
of cloud and has a maximum value $n\sim 10^6$ cm$^{-3}$. We have chosen
this limit because the gas will go through a rapid phase of transition
at density $n \geq 10^6$ cm$^{-3}$ via three-body $\rm H_2$ formation 
reaction \citep{pss83}} 
\begin{equation}
\rm H + H + H \rightarrow H_2 + H
\end{equation}
\begin{equation}
\rm H + H + H_2 \rightarrow H_2 + H_2.
\end{equation}
\noindent
{\bf We can therefore simulate the crucial transformation of the 
atomic to molecular hydrogen as the collapse proceeds to higher density.
The rapid formation of the molecular hydrogen cools the gas very 
efficiently and initiate chemothermal instability \citep{dutta15a}. 
Although the rate coefficient for this reaction is highly uncertain,
we choose the intermediate one that is 
$7.7 \times 10^{-31} T^{-0.464}$ cm$^6$ s$^{-1}$ \citep{g08}.}
The CH1 and CH2 contains gas mass of 1030 $M\odot$ and 1093 $M\odot$ 
respectively. {\bf The temperature at the center of both minihalos
has a maximum value $\sim 450 K$, while at the periphery it is around
$50 K$.} With altogether six million SPH particles in each of our 
simulations, we are able to resolve the angular velocity of the gas 
significantly better than in the SPH models of our previous study. 
The numerical resolution of our simulations for both minihalos are 
of the order of 0.01 $M\odot$. {\bf The initial rotation of minihalos,
defined by the spin parameter ($\beta_0$), is 0.035 (for CH1) and
0.042 (for CH2) respectively. So the realistic minihalos falls 
intermediate compared to the idealized gas clump where we simulate
the gravitational collapse of solid body with $\beta_0$ = 0.0, 0.005, 
0.01, 0.05, 0.1 and 0.2 respectively.}

{\bf The chemical network for primordial collapse is extremely complex
with primordial hydrogen, helium and deuterium mixing in a different
way to contribute toward the thermal, chemical and dynamical evolution
of gas. We follow \citet{dnck15} to model the time-dependent chemical 
reactions in this study. We use the modified version of the Gadget-2
momentum equation from our previous study \citet{dutta15b} to model
external pressure boundary for both the idealized as well as realistic
minihalos.}


%
%
%
%
%
%
%
%
%
%
%

\section{Angular Momentum}
\label{sec:angular}

\begin{figure*}
\centerline{
\includegraphics[width=5.3in]{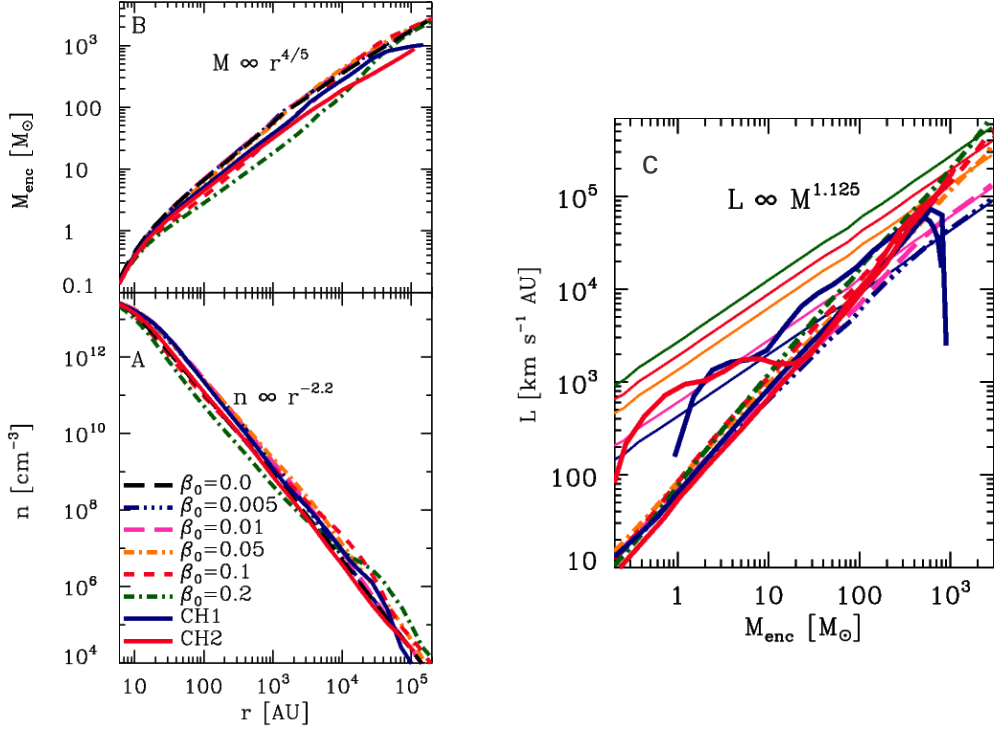}
}
\caption{\label{angmom} (A) Radial logarithmic binned, mass-weighted 
averages of the number density ($n$) and (B) enclosed gas mass within 
the radius ($r$) are plotted as a function of the radius for different 
degrees of initial rotation of the cloud ($\beta_0$), including for 
the non-rotating case ($\beta_0$ = 0). These plots are created just 
before the formation of the first protostar. The specific angular 
momentum follows a power-law relation with the enclosed gas mass 
irrespective of the cloud's initial configuration. The solid lines 
represent the initial state of the angular momentum ($L_0$) for the 
different $\beta_0$-modeled.
} 
\end{figure*}

\begin{figure*}
\centerline{
\includegraphics[width=5.0in]{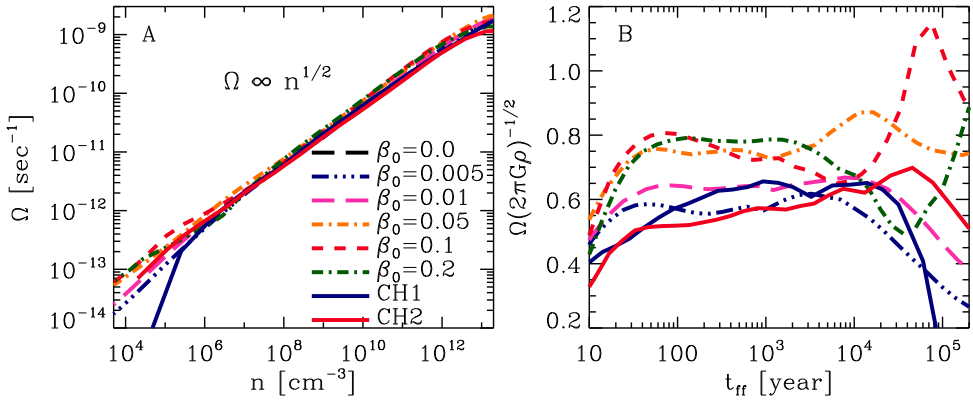}
           }
\caption{\label{omega} (A) Radially binned, mass-weighted averages 
of the angular velocity is plotted a function of the number density
($\Omega \propto n^{1/2}$) for different degrees of initial rotation
($\beta_0$). 
(B) Toomre's Q-parameter (defined as $Q = \Omega\sqrt{2\pi G\rho}$ 
(\citealt{mhn97}) is nearly constant with the local free-fall time. 
} 
\end{figure*}

It is a common feature of all star formation calculations that the 
angular momentum of a single star or star-cluster is much lower than 
that of the original cloud in which the protostars form \cite[e.g.][] 
{bodenheimer95}. We follow the collapse of primordial gas and study 
the distribution of the angular momentum for different degrees of 
initial rotation ($\beta_0$).

Fig.~\ref{angmom} shows the physical conditions in the gas once the 
central region has collapsed to a density of $\sim 5 \times 10^{13}$ 
cm$^{-3}$. The panels show mass-weighted averages of the properties 
of individual SPH particles {\bf within radial} logarithmic bins. We 
find in Fig.~\ref{angmom}A that regardless of the initial rotation 
speed, all clouds closely follow a power-law density profile 
$n \propto r^{-2.2}$, {\bf implying} to a relation between the radius 
and enclosed mass $M \propto r^{4/5}$ (Fig.~\ref{angmom}B). However, 
these are only approximate relations and different degrees of the
rotationally support clouds lead to subtle deviations. For example, 
larger rotational support leads to a slightly steeper density gradient 
in the inner $10^{4}$ AU, so that the enclosed mass within a certain 
radius is larger for the clouds with smaller $\beta_0$ values. 

We plot the final and initial specific angular momentum profiles 
(i.e., angular momentum per unit mass shell) in Fig.~\ref{angmom}C 
as a function of {\bf the} enclosed gas mass. 
The solid lines represent the cloud's initial specific angular 
momentum ($L_0$), while the dot-dashed lines describe the final 
state. We see that the specific angular momentum changes significantly, 
resulting in its redistribution during the gravitational contraction. 
For the final state, we find that $L \propto M^{1.125}$ irrespective 
of the initial rotation of the cloud. We also note in 
Fig.~\ref{angmom}A that it is the $\beta_0 = 0.2$ case that has the 
lowest values of $n$ within the enclosed $10^{3} M_\odot$ and the 
highest values of $n$ outside this range at the evolutionary phase 
recorded here. That means cloud with $\beta_0 = 0.2$ has gone 
through an efficient phase of angular momentum distribution from 
the center to the periphery during the initial contraction.

This characteristic power-law behaviour has been reported in 
several studies \cite[e.g.,][]{abn02,yoha06,cgsgkb11} using 
cosmological initial conditions that are very different from our 
idealized initial conditions. We find the scaling relations of 
the characteristic angular momentum profile in two completely 
different gas distribution: (1) in idealized minihalos of five 
different simulations (with $\beta_0$-model) and (2) in two 
realistic cosmological simulations. This rigorous {\bf experiment 
suggests} that the power-law angular momentum distribution is a 
natural outcome of the collapsing gas, and may be {\bf a} fundamental 
and universal property of the gravitational collapse of the primordial
star-formation cloud. 

\subsection{Origin of the power-law slope}

Hydrodynamical calculations of the gravitational collapse 
\cite[e.g.][]{on98} have shown that all clouds converge towards 
a higher\hbox{-}density regime that roughly obeys the power-law 
relation $n \propto r^{-2.2}$. We find that the non-rotating 
clouds ($\beta_0$ = 0) also have the similar density profile like
their rotating counterparts. Previous studies
have shown that the collapse approaches a self-similar solution, 
well-known for the non-rotating case \citep{larson69}. Even for 
rotating clouds, it appears that the collapse still proceeds in 
a self-similar fashion \citep{mhn97}. From a dimensional analysis, 
one has $\Omega/\sqrt{G\rho} = \mbox{constant}$, where $\Omega$ 
is the angular velocity of the rotating cloud. Hence we have  
$\Omega \propto n^{1/2} \propto R^{-1.1}$ (Fig.~\ref{omega}A), 
followed by $\Omega \propto M^{-1.375}$. This leads to the angular 
momentum profile of the collapsed gas:
\begin{equation}
\boxed{L \propto \Omega r^{2} \propto M^{1.125}\, .}
\end{equation}
The relation in Fig.~\ref{omega}A is also satisfied for a 
homogeneous cloud in centrifugal equilibrium. Even in that case, 
$\Omega_K = v_K/r = \sqrt{GM/r^3} \simeq \sqrt{G\rho(r)} $ for 
a mean density $\rho(r)$ within the radius $r$, where $\Omega_K$ 
is the angular velocity for Keplerian rotation. In Fig.~\ref{omega}B, 
we show the Toomre Q-parameter defined as 
$Q = \Omega/\sqrt{2\pi G\rho}$ (\citealt{mhn97}) 
plotted as a function of the free-fall time. As expected, the 
quantity $\Omega(2\pi G\rho)^{-1/2}$ is roughly constant with the 
free-fall time of the collapsed gas.
We conclude that the angular velocity and the angular momentum 
follow a pivotal power-law {\bf relation} with the enclosed 
mass and are independent of the initial rotation and turbulence 
of the cloud. At this point, we also note that the specific 
angular momentum for the final state of the collapse does not 
necessarily depend on the initial density configuration. As the 
collapse proceeds to protostellar densities, one can always 
obtain the power-law angular momentum profile for both the 
initial uniform density and centrally condensed distribution.

\subsection{Discussion on angular momentum distribution}

As the gas continues to collapse, the angular momentum is 
continually distributed to maintain the angular velocities 
that are significant fraction of the Keplerian velocities. 
Unlike the present-day star formation where the rotation 
rate is reduced by the magnetized stellar winds and magnetic 
braking \citep{mp10}, the process of transferring the angular 
momentum during the primordial collapse is ambiguous and of 
utmost interest. Since the initial strength of rotation of 
the clouds can significantly affect the thermal as well as 
dynamical evolution of gas, a higher angular momentum {\bf of} 
clouds can have different properties than their slowly 
rotating counterparts \citep{hirano14,dutta15b}.  

We adopt a Lagrangian point of view to calculate the time
derivative of the specific angular momentum ($\tau = dL/dt$) 
for different mass shells that are logarithmic binned 
mass-weighted averages of SPH particles. {\bf Following 
\citet{yoha06}}, we can write the expression for two kinds of 
torques in the case of primordial gas collapse, 

\begin{eqnarray}
\frac{dL}{dt} &=& \frac{d}{dt}\sum\limits_{i=1}^n \left(\vec{r}_i \times \vec{v}_i\right) \nonumber\\ &=& \sum\limits_{i=1}^n \vec{r}_i \times \left(\vec{F}_{\rm grav} + \vec{F}_{\rm pres}\right) \nonumber\\ &=& \vec{\tau}_{\rm grav} + \vec{\tau}_{\rm pres}
\end{eqnarray}

\noindent
{\bf where $\vec{F}_{\rm grav}$ is the total acceleration due 
to gravity, i.e., summation over all mass shell $i$, and 
$\vec{F}_{\rm pres}$ is the total force per unit mass due to 
the pressure gradients between the mass shells. We can therefore 
write the gravitational torques ($\tau_{\rm grav}$) and pressure 
torque ($\tau_{\rm pres}$) as following:}

\noindent
\begin{eqnarray}
\vec{\tau}_{\rm grav} = \sum\limits_{i=1}^n \vec{r}_i \times \frac{d\vec{v}_i}{dt} \nonumber \\ \vec{\tau}_{\rm pres} = \sum\limits_{i=1}^n \vec{r}_i \times \frac{\vec{\nabla P_i}}{\rho_i} 
\end{eqnarray}

\noindent
{\bf where $\vec{r}_i$ is the distance of the mass shell $i$ from 
the center, $d\vec{v}_i/dt$ is the acceleration due to gravity 
of the $i^{\rm th}$ shell, $\rho_i$ is the density and $\nabla P_i$ 
is the pressure gradient between the shells. The sum goes over all
SPH particles in spherical mass shell under consideration.}

The gravitational torque is generated due to the nonaxisymmetric 
nature of the collapsing cloud or due to the spiral structure, 
which is consistant with the findings of previous studies 
\cite[e.g.,][]{yoh08,lkk11}. {\bf For a rotating collapsing core, 
the non-radial gravitational forces due to the irregularities in 
the structure can produce a trailing spiral features of large 
amplitude \citep{larson84}}. The associated gravitational torques 
can then transfer angular momentum outward on an orbital timescale. 
In the case of a clustered formation, the angular momentum could 
be efficiently transported outward by tidal torques, similar to 
the case of present-day star formation in a clustered environment 
\citep{bbb03}. {\bf Gravitational torques associated with gravitational 
instabilities also become important for angular momentum transport 
and energy dissipation in a self-gravitating accretion disk within 
a local viscous framework \citep{lr04}.}
However, for a rotating collapsing core we find that the density 
gradient between the shells can produce the non-radial pressure 
torque. The concept of transferring angular momentum due to the 
pressure torque may be equivalent with the transport of angular 
momentum via hydrodynamic shocks during the turbulent collapse 
\cite[as discussed in][]{abn02}.

\begin{figure}
\centerline{
\includegraphics[width=3.2in]{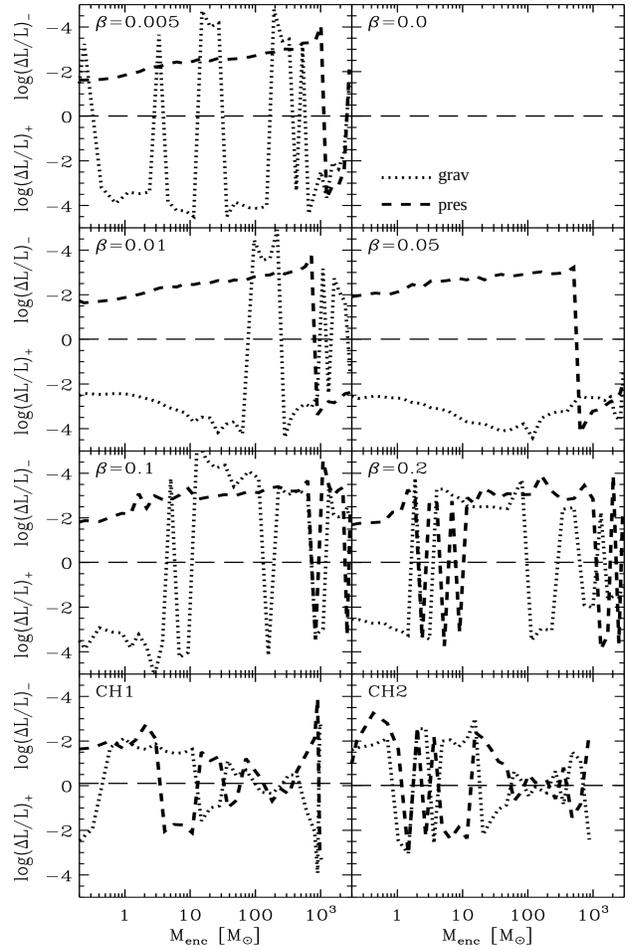}
           }
\caption{\label{fracL} The fractional change of the specific angular 
momentum ($\Delta L/L$) with the free-fall time of the mass shell 
due to the torques acting on the shell are plotted as a function of 
enclosed gas mass, just before formation of first protostar. The sign 
implies whether the angular momentum is gained (positive) or lost 
(negative) during collapse.
} 
\end{figure}

{\bf For all $\beta_0$ modeled, we take into account both kind of 
forces in our simulations: one is the acceleration due to gravity 
that leads to the gravitational torque ($\tau_{\rm grav}$) and another
is the pressure-torque ($\tau_{\rm pres}$). Here we refer the study 
by \citet{larson84} that concluded that the non-axisymmetric nature 
of the cloud or inhomogeneities tend to form a spiral structure that 
produces the gravitational torque. The pressure forces can arise from 
the density gradient between the mass shells. We have observed that 
the gravitational and pressure torque act in opposite direction, 
though not always, and change direction throughout the collapsing
gas cloud. However, the contribution of the torques will be more clear 
in terms of change of angular momentum that is shown in Fig.~\ref{fracL}.} 

{\bf Taking the fractional change in the angular momentum of a shell
($\Delta L(r)/L(r)$) with the local free-fall time ($t_{\rm ff}$), 
we find that both the torques are dominant throughout the evolution. 
As the collapse proceeds to higher densities, the angular momentum 
of a shell can increase or decrease depending on whether torques 
acting on the shell are positive or negative. We use $(\Delta L/L)_+$ 
and $(\Delta L/L)_-$ for the positive and negative fractional change 
of each type of torque (Fig.~\ref{fracL}). The net change in the 
angular momentum happens due to the synchronous action of both the 
gravitational and pressure torque. However, in the inner regime, the 
cloud gains angular momentum due to the pressure gradient between 
the mass shell. On the other hand cloud's angular momentum is lost 
due to torque exerted by the gravity.}
This is consistent with the recent work by \citet{gbcgskys12} that 
concluded that the gravitational torque tends to decreases the angular 
momentum whereas the pressure torque does the opposite during the 
collapse of gas towards the protostellar densities. The spiral structure
of the Keplerian disk that becomes gravitationally unstable can
transfer the mass inside, thereby removing the extra angular momentum 
in the outward direction \citep{bgsh15}. In addition, cosmological 
hydrodynamics simulations by \citet{wta08} has shown that the 
rotational secular bar instabilities can efficiently transport 
angular momentum outward in the central regime. It has also been 
suggested that for a simple hydrodynamical flow, the Reynolds stress 
associated with velocity fluctuations might act as a turbulent 
transport, and angular momentum can then be removed from the mean 
flow \citep{lodato08}.

\section{Summary and discussion}
\label{sec:summary}

We have investigated in detail the angular momentum evolution 
during the collapse of the primordial gas and analysed its
relation with other physical quantities. For this purpose, we 
have followed a systematic parameter study that spans two 
orders of magnitude in the strength of rotational support for 
the idealized solid body rotation along with two realistic 
cosmological minihalos.

We find that the initial configuration of the gas clouds (e.g. 
rotation, turbulence and halo-to-halo variation) have little 
influence on the distribution of the cloud's specific angular 
momentum. Apart from the well-known density distribution, we 
find that the angular momentum {\bf also provides} another way 
of explaining the properties of the primordial gas collapse. 
In addition, the sole dependency of the angular momentum on 
the gas mass makes it easier to understand the primordial star 
formation. These basic characteristics of the angular momentum 
distribution make it to be the fundamental property of the 
primordial gas collapse. We also discuss the possible mechanisms 
that are responsible for the power-law profile of angular momentum,
mainly the simultaneous effect of the gravitational and pressure 
{\bf torques}. We argue that the distribution is due to internal 
torques only and does not require the presence of turbulence or 
magnetic fields. However, the exact way of transferring the 
angular momentum, and {\bf whether or not} it is related to the 
other physical properties, need further theoretical investigation. 

{\bf The angular momentum distribution is an area of active research.
There have been various investigations in a number of previous studies 
for the transport issues. For example, from the forefront studies by 
\citet{larson69} to present theoretical studies \cite[e.g.,][]{yoha06,
b13,hirano15}, people have tried to describe the plausible causes of 
the angular momentum transport especially for a system where magnetic 
fields have less or no contribution. The gravitational collapse of 
primordial gas is considered to be weakly magnetized (although 
recent simulations are showing the influence of the magnetic field,
e.g., \citealt{sur10}). In such system, the pressure torque and 
gravitational torque could play an important role (although some 
studies consider only gravitational torque). We have performed 
a set of high-resolution simulations with varied rotation parameter 
($\beta_0$) to study the angular momentum distribution. For the 
first time, to the best our knowledge, we have shown the effect of 
both the pressure torque as well as gravitational torque. Although 
our approach is different, we obtain similar results from previous 
forefront investigations \cite[e.g.,][]{yoha06,gbcgskys12}. Our 
primary aim in this study was to investigate the distribution of 
the angular momentum, and we have shown how the distribution is
related to the physical parameter of the collapsing gas. However, 
in describing the distribution along with the scaling relation, 
we have also made an attempt to discuss how the power-law 
behavior is controlled during the collapse. From our investigation, 
we can conclude that the distribution of angular momentum happens 
due to the concurrent effects of internal torques in a collapsing 
system where turbulence or magnetic field does not play any 
significant role.}

{\bf At this point, we would like to emphasize that we have discussed 
the possible source of transport only for a weakly magnetised 
Pop~III collapse \citep{tm04,byhm09}. On the other hand, the study 
by \citet{deSouza08} has shown that the magnetic field plays an 
important role during collapse. The most common explanation for 
transfer of angular momentum is magnetized stellar winds and 
magnetic outflow (see e.g.\ \citealt{mp05}). It has also been 
suggested that accretion disks threaded by a weak magnetic field are 
subject to magneto-hydrodynamics (MHD) instabilities \citep{bh98}, 
which can induce turbulence in the disk, thereby being able 
to transport angular momentum and to promote the accretion process. 
However, in many astrophysical phenomena, such as the outer 
regions of protostellar disks, the ionization level is expected 
to be low, {\bf significantly reducing} the effects of magnetic 
fields in determining the dynamics of the disk \citep{gammie96}.}

{\bf More recently the numerical simulations show the amplification 
of small seed fields up to dynamically important strengths during 
the collapse of primordial star-forming halos 
\citep{sur10,sur12,sbsakbs10,ssfgkb12}. The angular momentum can 
then be transported in the protostellar disk via magnetic braking 
\citep{md13}. However, the influence of the amplified magnetic 
fields on the redistribution of angular momentum requires 
further investigation.}

\smallskip
The author acknowledges Prateek Sharma, Biman Nath and Kazu Omukai 
for the helpful comments on the manuscript. {\bf The author also wish 
to thank the referee for constructive suggestions that have helped 
to improve the quality of the paper.} The author is grateful to the 
Centre for Theoretical Studies (CTS) of the Indian Institute of 
Technology $\hbox{--}$Kharagpur and the Department of Physics of the
Indian Institute of Technology$\hbox{--}$Banaras Hindu University, 
Varanasi for the local hospitality. {\bf The author would also like 
to thank the Inter-University Center for Astronomy and Astrophysics 
(IUCAA) at Pune for the local hospitality and financial support. The 
present work is supported by the Indian Space Research Organization 
(ISRO) grant (No.~ISRO/RES/2/367/10-11).} \\

\footnotesize{

}

\end{document}